\input epsf
\input harvmac.tex

\def\capt#1{\narrower{
\baselineskip=14pt plus 1pt minus 1pt #1}}

\lref\sas{Zamolodchikov, A.B.: Fractional-spin integrals of
motion in perturbed conformal field theory.
In: Guo, H., Qiu, Z. and Tye, H. (eds.)
Fields, strings and quantum gravity. Proceedings,
Beijing 1989. New York: Gordon and Breach science publishers,
1990}

\lref\BPZ{Belavin, A.A., Polyakov, A.M. and
Zamolodchikov, A.B.:
Infinite conformal symmetry in two-dimensional
quantum field theory.
Nucl. Phys. {\bf B241}, 333-380 (1984)}

\lref\pokr{Patashinskii, A.Z. and Pokrovskii, V.N.:
Fluctuation theory of phase transition,
Oxford: Pergamon Press 1979}

\lref\carmus{L\'assig, M., Mussardo, G. and Cardy, J.L.:
The scaling region of the tricritical Ising model
in two-dimensions. Nucl. Phys. {\bf B348}, 591-618 (1991)}

\lref\shifm{Shifman, M.A., Vainshtein, A.I. and Zakharov, V.I.:
QCD and resonance physics. Theoretical foundation.
Nucl. Phys. {\bf B147}, 385-447 (1979)\semi
QCD and resonance physics. Applications.
Nucl. Phys. {\bf B147}, 448-534 (1979)} 

\lref\zamole{Zamolodchikov, A.B.: S-matrix of the 
sub-leading magnetic perturbation of the tricritical Ising,
PUPT 1195-90}

\lref\ReSm{Reshetikhin, N.Yu. and Smirnov, F.A.:
Hidden quantum group symmetry and
integrable perturbations of conformal
field theories. Commun. Math. Phys.
{\bf 131}, 157-178 (1990)}

\lref\BeLeC{Bernard, D. and LeClair, A.:
Residual quantum symmetries of
the restricted Sine-Gordon theories.
Nucl. Phys. {\bf B340},
721-751 (1990)}

\lref\Yur{Yurov, V.P. and Zamolodchikov, Al.B.:
Truncated conformal space approach to scaling Lee-Yang model.
Int. J. Mod. Phys. {\bf A5}, 3221-3245 (1990)}

\lref\Sm{Smirnov, F.A.: Reductions of Quantum Sine-Gordon Model
as Perturbations of Minimal Models of Conformal Field Theory.
Nucl. Phys. {\bf B337}, 156-180 (1990)}

\lref\Leclair{LeClair, A.: Restricted Sine-Gordon theory
and the minimal conformal series. Phys. Lett. {\bf B230}, 103-107 (1989)} 

\lref\WuM{Wu, T.T., McCoy, B.M., Tracy, C.A. and
Barouch, E.: Spin-spin correlation functions for the
two-dimensional Ising model: Exact theory in
the scaling region. Phys. Rev.  {\bf B 13}, 1, 316-374
(1976)}

\lref\mag{Guida, R. and Magnoli, N.: Tricritical
Ising Model near criticality,
Preprint SPhT-t96/142, GEF-Th-15,
\#hepth  961254}

\lref\acerbi{Acerbi, C.: Form factors of 
exponential operators and wave
function renormalization constant in the  Bullough-Dodd
model. Nucl. Phys. {\bf B497}, 589-610 (1997)}

\lref\Zar{Zamolodchikov, Al.B.: Thermodynamic Bethe ansatz in
relativistic models: Scaling 3-state Potts and Lee-Yang models.
Nucl. Phys. {\bf B342}, 695-720 (1990)}

\lref\brl{Brazhnikov, V. and Lukyanov, S.: Angular quantization and
form-factors in massive integrable models,
Preprint RU-97-58, CLNS 97/1488, $\#$hep-th/9707091}

\lref\Arien{Arinshtein, A.E., Fateev, V.A.
and Zamolodchikov, A.B.:
Quantum S matrix of the\ $(1+1)$-dimensional Todd chain.
Phys. Lett. {\bf B87}, 389-392 (1979)}

\lref\Baxter{Baxter, R.J.: Exactly solved models
in statistical mechanics. London: Academic Press 1982}

\lref\BF{Forrester, J. and Baxter R.: The eight-vertex SOS model and the
Rogers-Ramunujan identities. J. Stat. Phys.
{\bf 38}, 435-472 (1985)}

\lref\ZZ{Zamolodchikov, A.B. and Zamolodchikov, Al.B.:
Structure Constants and Conformal Bootstrap in
Liouville Field Theory.
Nucl. Phys. {\bf B477}, 577-605   (1996) }

\lref\Zarn{Zamolodchikov, Al.B.: Mass scale
in the Sine-Gordon model and its
reductions. Int. J. Mod. Phys. {\bf A10}, 1125-1150  (1995) }

\lref\mus{Koubek, A., Martins, M.J.. and Mussardo G.:
Scattering matrices for $\Phi_{(1,2)}$ perturbed conformal
minimal model in absence of kink states.
Nucl. Phys. {B368}, 591-610 (1992)}

\lref\martins{Martins, M.J.: Constructing a S matrix
from the truncated conformal approach data.
Phys. Lett. {\bf B262}, 39-41 (1991)}

\lref\kitaev{Kitaev, A.V.: Method of isomonodromic 
deformations for ``degenerate'' third Painleve equation.
J. Sov. Math. {\bf 46}, 2077-2083 (1989)}

\lref\tracy{Tracy, C.A. and Widom, H.:
On exact solutions to the cylindrical
Poisson-Boltzmann equation with applications to
polyelectrolytes.  Preprint 1997, $\#$cond-mat/9701067}

\lref\takacs{Kausch, H., Takacs, G. 
and Watts, G.: On the relation between
$\Phi_{(1,2)}$ and $\Phi_{(1,5)}$ perturbed minimal models and unitarity.
Nucl. Phys. {\bf B489}, 557-579 (1997)}

\lref\zamola{Zamolodchikov, Al.B.: Mass scale
in the Sine-Gordon model and its
reductions. Int. J. Mod. Phys. {\bf A10}, 1125-1150  (1995) }

\lref\ZamAl{
Zamolodchikov, Al.B.:
Two-point correlation function
in Scaling Lee-Yang model.
Nucl. Phys. {\bf B348}, 619-641 (1991)}

\lref\LZ{Lukyanov, S. and Zamolodchikov, A.:
Exact expectation values of local fields
in quantum sine-Gordon model.
Nucl. Phys. {\bf B493}, 571-587 (1997)}

\lref\FLZZ{Fateev, V., Lukyanov, S., Zamolodchikov, A. and
Zamolodchikov, Al.: Expectation values of boundary fields
in the boundary sine-Gordon model. Phys. Lett. {\bf B406}, 
83-88 (1997)}

\lref\fatt{Fateev, V.A.: The exact relations between the
Coupling constants and the masses of particles for the
integrable perturbed conformal field theories.
Phys. Lett. {\bf B324}, 45-51 (1994)}

\lref\magnol{Guida, R. and Magnoli, N.: Vacuum expectation values from
a variational approach. Preprint SPhT-t97/055, GEF-Th-5,
\#hepth  9706017}

\lref\Dod{Dodd, R.K. and Bullough, R.K.: Polynomial
conserved densities for the sine-Gordon equations.
Proc. Roy. Soc. Lond. {\bf A352}, 481-502 (1977)}

\lref\Zhib{Zhiber, A.V. and Shabat, A.B.: Klein-Gordon
equations with a nontrivial group. Sov. Phys. Dokl.
{\bf 24}, 607-609 (1979)}

\lref\smir{Smirnov, F.A.: Exact S-matrices for\
$\Phi_{1,2}$-perturbed minimal models of conformal field theory.
Int. J. Mod. Phys. {\bf A6}, 1407-1428 (1991)}

\Title{\vbox{\baselineskip12pt
\hbox{CLNS 97/1510}
\hbox{RU-97-74}
\hbox{hep-th/9709034}}}
{\vbox{\centerline{
Expectation values of local fields}
\vskip6pt
\centerline{in Bullough-Dodd model and}
\vskip6pt
\centerline{  integrable
perturbed conformal field theories}}}

\centerline{Vladimir Fateev$^{1,4}$,
Sergei Lukyanov$^{2,4}$,}
\centerline{
Alexander Zamolodchikov$^{3,4}$
and Alexei Zamolodchikov$^{1}$}

\centerline{}
\centerline{$^1$Laboratoire de Physique
Math\'ematique, Universit\'e de Montpellier II}
\centerline{ Pl. E. Bataillon,  34095 Montpellier, FRANCE}
\centerline{$^2$Newman Laboratory, Cornell University}
\centerline{ Ithaca, NY 14853-5001, USA}
\centerline{$^3 $Department of Physics and Astronomy,
Rutgers University}
\centerline{ Piscataway,
NJ 08855-0849, USA}
\centerline{and}
\centerline{$^4$L.D. Landau Institute for Theoretical Physics,}
\centerline{Chernogolovka, 142432, RUSSIA}

\centerline{}
\centerline{}
\centerline{}

\Date{August, 97}

\vfill
\eject

\centerline{\bf Abstract}

\bigskip
Exact expectation values of the fields $e^{a\varphi}$ in the
Bullough-Dodd model are derived by adopting the ``reflection
relations'' which
involve the reflection S-matrix of the Liouville theory, as well as
special analyticity assumption. Using this result we propose 
explicit expressions for  expectation values of all primary operators
in the $c<1$ minimal CFT perturbed by the operator $\Phi_{1,2}$
or $\Phi_{2,1}$. Some
results concerning the $\Phi_{1,5}$
perturbed minimal models are also presented.

\vfill
\eject

\newsec{Introduction}

\vskip 0.1in

Computation of vacuum  expectation values (VEV)
of local fields (or one-point correlation functions) is important
problem of quantum field theory (QFT)\ \pokr,\ \shifm.
When applied to statistical mechanics the VEV determine ``generalized
susceptibilities'', i.e. linear response of the system to external
fields. More importantly, in QFT defined as a perturbed conformal field
theory the VEV provide all information about its correlation functions
which is not accessible through straightforward calculations in conformal
perturbation theory\ \ZamAl.
Recently some progress was made in calculation
of the VEV in 1+1 dimensional
integrable QFT. In\ \LZ\ explicit expression for
the VEV of exponential fields 
in the  sine-Gordon and sinh-Gordon models was
proposed. It was found in\ \FLZZ\ that this expression can be obtained as
minimal solution to certain ``reflection relations'' which involve the
Liouville ``reflection S-matrix''\ \ZZ,
provided one assumes simple analytic
properties of the VEV. This result for the sine-Gordon model allows one to
obtain, through the quantum group restriction, expectation values
of primary fields in $c<1$ minimal CFT perturbed by the operator
$\Phi_{1,3}$, with good agreement with numerical data\ \magnol.
In this paper
we use the ``reflection relations'' to obtain the VEV of exponential fields
$e^{a\varphi}$ in the   so called Bullough-Dodd model\ \Dod, \Zhib.
As is known\ \smir, for special pure imaginary values of its coupling
constant the Bullough-Dodd model
admits quantum group restriction leading to a $c<1$ minimal conformal
field theories (CFT)
perturbed by the operator $\Phi_{1,2}$. We use this relation to obtain
the VEV of primary fields in these perturbed minimal CFT.

In Sect.2 we present some details of the derivation of the VEV in
the sinh-Gordon
and sine-Gordon models using the ``reflection relations'', and show how
the VEV of primary fields in minimal CFT perturbed by $\Phi_{1,3}$
can be obtained. In Sect.3 we extend this approach and find explicit
expression for the VEV of the exponential fields $e^{a\varphi}$ in the
Bullough-Dodd model. We show that in the semi-classical limit our
expression agrees with known results from the
classical Bullough-Dodd
theory. We also run some perturbative checks. In Sect.4 we study the
minimal CFT perturbed by the operator $\Phi_{1,2}$. Using our result for
the Bullough-Dodd model we 
propose exact formula for the VEV of all primary
fields $\Phi_{l,k}$ in these perturbed theories. We also compare our
results with numerical 
data available in literature. In Sects.5 and 6 some
results and conjectures concerning minimal models perturbed by the
operators $\Phi_{1,5}$ and $\Phi_{2,1}$ are presented.

\vskip 0.2in

\newsec{Reflection relations in the sinh-Gordon model}

\vskip 0.1in

The sinh-Gordon model is defined by the Euclidean action
\eqn\shg{
{\cal A}_{shG}=\int d^2 x\,
\Big\{\, {1\over{16\pi}}(\partial_{\nu}\varphi)^2 + \mu e^{b\varphi} +
\mu e^{-b\varphi}\, \Big\}\ . }
We are interested in the expectation values of
the exponential fields,
\eqn\shgexp{ G(a)=\langle\, e^{a\varphi}\, \rangle_{shG}\ .}
As was observed in\ \FLZZ\ these 
expectation values satisfy the ``reflection
relation''
\eqn\shgra{ G(a)=R(a)\,   G(Q-a)\, ,}
\eqn\shgrb{G(-a) =  R(a)\,  G(-Q+a)\, ,}
where 
\eqn\Qb{Q=b^{-1}+b}
and the function $R$ is related to the so called Liouville
reflection amplitude $S$\ \ZZ,
\eqn\sliouville{
R\big({Q\over 2}+iP\big) =S(P)=
-\bigg({{\pi\mu\, \Gamma(b^2)}\over{\Gamma(1-b^2)}}\bigg)^{-{2 iP\over b}}\
{{\Gamma\big(1+2iP/b\big)\,
\Gamma\big(1+2iPb\big)}\over{\Gamma\big(1-2iP/b\big)\,
\Gamma\big(1-2iPb\big)}}\ . }
Note that the second of the relations\ \shgrb\ follows from the first 
one if one
takes into account an obvious symmetry of\ \shgexp,
$$
G(a)=G(-a)\ .
$$
No rigorous proof of the reflection 
relations\ \shgra, \shgrb\ is known to us. 
Here we give simple intuitive argument in support of these relations.

Let us note that the sinh-Gordon theory\ \shg\ can be interpreted as the
perturbed Liouville QFT in two different ways. First, one could take the
first two terms
in the action\ \shg\ as the 
action\ ${\cal A}_{L}$\ of the Liouville theory (in a
flat 2D background metric) and treat the last term containing
$e^{-b\varphi}$ as the perturbation. Then naively one could write down
the conformal perturbation theory series (expansion in the
perturbation term) for the one-point function of\ \shg,
\eqn\cpt{\langle\,  e^{a\varphi}(x)\, \rangle_{shG}= 
Z^{-1}\sum_{n=0}^{\infty}{{(-\mu)^n}\over n!}\int d^2 y_1 \ldots d^2 y_n
\, \langle\,  e^{a\varphi}(x)
e^{-b\varphi}(y_1) \ldots e^{-b\varphi}(y_n)\, \rangle_{L}\ ,}
where $\langle\, \ldots\, \rangle_{L}$\  are
the expectation values over the Lioville theory\ ${\cal A}_{L}$,
and $Z$ is its partition function which does not depend on $a$.
With this expression the 
first ``reflection relation''\ \shgra\ follows from the
reflection property of the Liouville correlation functions
(see\ \ZZ\ for
the details),
\eqn\lrp{\langle\,  e^{a\varphi}(x)\ldots\, \rangle_{L} =
R(a)\  
\langle\,   e^{(Q-a)\varphi}(x)\ldots\, \rangle_{L}\ ,}
where dots stand for any local insertions.
The coefficient function $ R(a)$ is
related to the Liouville two-point correlation function
\eqn\twol{\langle\,  e^{a\varphi}(x)e^{a'\varphi}(x')\, \rangle_{L} =
\big[\, \delta_{Q-a,a'}+ R(a)\,\delta_{a, a'}\, 
\big]\  |x-x'|^{-4a (Q-a)}\ ;}
its explicit form is given by\ \sliouville. The
function $S$ in\ \sliouville\
can be interpreted as the amplitude of scattering off the ``Liouville
wall'', as explained in\ \ZZ. Alternatively, one could interpret the
second term in\ \shg\ as the perturbation of 
the Liouville CFT defined by
the first and the third terms in\ \shg. Then writing down corresponding
naive conformal perturbation theory
series analogous to\ \cpt\ one would arrive at the second
relation\ \shgrb. In both cases the problem is that the integrals in
\cpt\ (as well as the integrals appearing with the second
interpretation) are highly infrared divergent and therefore the naive
series\ \cpt\ does not give a viable definition of the one-point
function. 

One can get around the above infrared problem as follows. Consider 2D 
``world sheet'' $\Sigma_g$, topologically a sphere, equipped with the 
metric $g_{\nu\sigma}(x)$, and 
define a version of the 
sinh-Gordon theory on $\Sigma_g$ with the following
non-minimal coupling to the background metric $g$,
\eqn\shgg{
{\cal A}^{g}_{shG}=\int d^2 x \sqrt{g}\ 
\Big\{\, {1\over{16\pi}}\, g^{\nu\sigma}\partial_{\nu}\varphi 
\partial_{\sigma}\varphi + {Q {\hat R} \over 8\pi}\, \varphi\, + 
\mu :e^{b\varphi}:_{g} +\mu :e^{- b\varphi}:_{g}\, \Big\}\ ,}
where $Q$ is given by\ \Qb, ${\hat R}$ denotes the scalar curvature 
of $g$ and the symbol
$:e^{\pm b\varphi}:_{g}$ signifies that these exponential fields are 
renormalized with respect 
to the background metric $g$. The first three
terms in\ \shgg\ define conformaly invariant Liouville 
theory\ ${\cal A}^{g}_{L}$\ on
$\Sigma_g$ so that\ \shgg\ agrees with the first of the above
interpretations of the sinh-Gordon model as the perturbed Liouville
theory ${\cal A}_L$.
Precisely this was our reason for adding the curvature term 
in\ \shgg. Due to its conformal invariance the Liouville theory is 
insensitive to a choice of the background metric $g$. If one picks
a conformal coordinates on $\Sigma_g$, so that
\eqn\gconf{{g}_{\nu\sigma}(x)=\rho (x)\,\delta_{\nu\sigma}\ ,}  
the dependence on $\rho(x)$ can be expelled from the Liouville part of
the action\ \shgg\ by the shift
\eqn\shift{ \varphi(x) \to \varphi(x) - Q\log{\rho}(x)\ .}
This transformation brings\ \shgg\ to the form
(up to field independent constant) 
\eqn\shggg{{\cal A}_{shG}^{g}=\int d^2 x\,  
\Big\{\, {1\over{16\pi}}\, (\partial_{\nu}\varphi)^2 
 + \mu e^{b\varphi} + \mu{\rho}^{2+2b^2}e^{-b\varphi}\, \Big\} 
+ Q \varphi_{\infty}\ ,} 
where now the exponential fields 
$e^{\pm b\varphi}\equiv :e^{\pm b\varphi}:_{g^{(0)}}$ are normalized with 
respect to the flat metric $g_{\nu\sigma}^{(0)}=\delta_{\nu\sigma}$,
so that
$$
e^{\pm b\varphi}(x)=[\rho(x)]^{-b^2}:e^{\pm b\varphi}:_{g}\ .
$$
The term with
$\varphi_{\infty}=\lim_{|x|\to \infty}\varphi(x)$ plays no role
in the perturbed theory. To be definite, let us
take the metric $g$ to be that of a sphere with area  $A$,
\eqn\gsphere{\rho(x)= \big( 1+\pi |x|^2/A \big)^{-2}\ .}
For finite $A$ the conformal perturbation theory for
$\langle\, e^{a\varphi}\, \rangle^g_{shG}$ in\ \shggg\ (the expansion in
$e^{-b\varphi}$) makes much better sense because now the integrals
analogous to those in\ \cpt\ contain the factors
$\prod_{k=1}^{n}\big[\, \rho(y_k)\, \big]^{2+2b^2}$ providing
an efficient infrared
cutoff. As the result these calculations produce a power series
of the form
\eqn\shgser{
\langle\, e^{a\varphi}\, \rangle^g_{shG} = \mu^{-{{a}\over b}}\,
A^{a(a-Q)}\ \sum_{n=2}^{\infty}\, \big[\, \mu A^{1+b^2}\, 
\big]^{2 n}\ G_n (a)\ .}
Owing to the property\ \lrp\ of the Liouville correlation function each
term in\ \shgser\ satisfies 
the reflection relation\ \shgra. Assuming that
the series\ \shgser\ defines a function $G(a,t)=\sum_{n=2}^{\infty} t^n
\, G_n (a)$ with the asymptotic 
\eqn\jsudt{G(a,t)\to\, G(a)\ t^{{a\over 2b}(1-{a\over
Q})}\, \ \ \ \ \ \ \ \ {\rm as}\ \ t\to\infty }  
and taking the limit $A\to\infty$\ (which brings\ \shgg\ back
to\ \shg) one arrives at \shgra.
Similarly, starting with the action
which differs from\ \shgg\ only in the sign of the
curvature term, one can repeat the above arguments, this time taking the
term with $e^{-b\varphi}$ as the perturbation. This leads to \shgrb.

Of course, these arguments do not give a rigorous proof of the reflection
relation\ \shgra\ because the convergence of the conformal perturbation
theory\ \shgser\ is
not at all obvious even at finite $A$. Also, the existence of 
appropriate limiting behavior $t\to\infty$\ in\ \jsudt\ is at
least problematic.
On the other
hand if one {\it assumes} these
arguments valid the reflection relations\
\shgra,\ \shgrb\ can be taken as the starting point in {\it deriving} the
expectation values \shgexp. In fact, if nothing is said about analytic
properties of the function $ G(a)$ the
equations\ \shgra,\ \shgrb\ are not
nearly sufficient to determine it. However, if one makes additional
assumption that $G(a)$ is a {\it meromorphic} function of $a$,
the following ``minimal solution'' to the
equations\ \shgra,\ \shgrb\ is readily
derived
\eqn\shgev{
\eqalign{
\langle\,  e^{a\varphi}\,  \rangle_{shG}=&
\bigg[\,
{m\,\Gamma\big(
{1\over 2+2b^2}\big)\,
\Gamma\big(1+{b^2\over 2+2b^2}\big)\over 4\sqrt{\pi}}
\, \bigg]^{-2a^2}
\times\cr
&\exp\biggl\lbrace\int_{0}^{\infty}{{dt}\over t}
\bigg[\ - {{\sinh^2 ( 2ab t )}\over{2\sinh(b^2 t)\, \sinh(t)\, 
\cosh\big((1+b^2)t\big)}}+
2a^2\,e^{-2t}\ \bigg]\, \biggl\rbrace\ ,}}
where\ \Zarn 
\eqn\sjusy{
m={4\sqrt{\pi}\over \Gamma\big({1\over 2+2b^2}\big)\,
\Gamma\big(1+{b^2\over 2+2b^2}\big) }\ 
\bigg[ -{\mu \pi \Gamma\big(1+b^2\big)\over \Gamma\big(-b^2\big)}\, 
\bigg]^{{1\over 2+2 b^2}}   }
is the particle mass of the  sinh-Gordon model. 
Note that\ \shgev\ is exactly the expression conjectured in\ \LZ. At the
moment we have absolutely no clue on how to justify this analyticity
assumption. We can only make a remark that while the above arguments
leading to the reflection relations\ \shgra,\ \shgrb\ do not seem to
depend on the
integrability of the sinh-Gordon model, the simple analytic properties
assumed above most likely do\foot{This can be compared
with the situation in lattice models of statistical
mechanics. While so called ``inversion relations'' (see\ \Baxter ) for the
partition function can be written down for many models including 
non-integrable ones, only integrable lattice models provide enough 
analyticity for making the inversion relations  a powerful tool of
computing the partition functions.}. However, we consider various
perturbative checks of\ \shgev\ performed in\ \LZ\ as a strong evidence
supporting both the above arguments about the reflection relations and
the analyticity assumption. Additional support is provided by the
results in Ref.\ \FLZZ\  where these assumptions are used to derive the
expectation values of the boundary operators in boundary sine-Gordon
model with zero bulk mass.
Furthermore, in Sect.3
we will use the same assumptions to obtain the expectation values of
exponential fields in the  Bullough-Dodd model.

As mentioned in\ \LZ,\ the expression\ \shgev\ can be used to obtain the
expectation values $\langle\,  \Phi_{l,k}\, \rangle$ of primary fields
with conformal dimensions
\eqn\hsydtr{\Delta_{l,k}={(p'l-pk)^2-(p'-p)^2\over 4 p p'} }
in 
perturbed ``minimal models''\ \BPZ 
\eqn\mmp{{\cal M}_{p/p'}+\lambda\int d^2 x \,\Phi_{1,3}(x)\ . }
This is possible because the perturbed minimal
models\ \mmp\ can be understood in terms of 
``quantum group restriction'' of the sine-Gordon
model\ \Leclair,\ \Sm,\ \ReSm,\ \BeLeC
\eqn\sg{{\cal A}_{sG} = \int d^2 x
\, \Big\{\, {1\over{16\pi}}(\partial_{\nu}\varphi)^2 - 
2\mu \cos{\beta\varphi}\, \Big\}\ .} 
As is known the sine-Gordon theory in infinite space-time exhibits a
symmetry with respect to affine quantum group $U_{q}({\hat{sl}_2})$
with the level equal to zero and 
\eqn\qdef{q=e^{{i\pi\over{\beta^2}}}\ .}
The soliton-antisoliton doublet transforms as
two-dimensional irreducible representation while the bound states
are scalars. The S-matrix commutes with the generators $E_{\pm},
H_{\pm}, F_{\pm}$ which satisfy the relations
\eqn\com{[E_{+}, F_{+}]={{q^{H_{+}}-q^{-H_{+}}}\over{q-q^{-1}}}\, ; \quad 
[E_{-}, F_{-}]={{q^{H_{-}}-q^{-H_{-}}}\over{q-q^{-1}}}\, ; \quad 
H_{+} + H_{-} =0\ .}
The operator $H_{+}$ is identified with the soliton charge. Important
observation made in\ \ReSm,\ \BeLeC\ is that special exponential fields
\eqn\vdef{V_{1,k}(x) =
e^{i{1-k\over 2}\beta\varphi}(x) \ \ \ \ (\, k=1,2,...\, ) }
commute with the generators $E_{+}, H_{+},
F_{+}$ of the subalgebra $U_q (sl_2)_{+}\in U_{q}({\hat{sl}_2})$,
\eqn\vcomm{
[E_{+}, V_{1,k}(x)]=[H_{+}, V_{1,k}(x)]=[F_{+}, V_{1,k}(x)]=0\ .}
This subalgebra plays the central role in the relation between\ \sg\ and
\mmp. The space of states ${\cal H}_{sG}$ of the  sine-Gordon model admits 
special inner
product\ \ReSm\ (different from the standard sine-Gordon scalar product) 
such that $E_{+}^{\dagger}=F_{+}$ (the standard scalar product implies
$E_{+}^{\dagger} = F_{-}$). If $q$ is a root of $1$, i.e.
\eqn\betap{\beta^2 = {{p}\over p'}\ ,}
where $p, p'$ are relatively prime  integers such that $p' > p>1$, one can
isolate the subspace ${\cal H}_{p}\in {\cal H}_{sG}$ consisting of
the representations of $U_q (sl_2)_{+}$ with the spins\ $j=0, 1/2, 1,
\ldots , p/2-1$.
The space $Inv\big({\cal H}_{p}\big)$ of 
invariant tensors of ${\cal H}_{p}$
is identified with the space of states of the
perturbed minimal model\ \mmp. The relation between the sine-Gordon
parameter $\mu$ and the coupling constant $\lambda$ in\ \mmp\ is
found in\ \Zarn,
\eqn\ksjdtr{\lambda  ={\pi\ \mu^2 \over (1-2\beta^2)(3\beta^2-1)}\ 
\biggl[\, {\Gamma^3(1-\beta^2)\, \Gamma(3\beta^2)\over 
\Gamma^3(\beta^2)\, \Gamma(1-3\beta^2) }\, \biggr]^{{1\over 2}}\ .}
The theory\ \mmp\ has $p-1$ degenerate ground states
$\mid 0_s \rangle,\  s=1, 2, \ldots , p-1$\ \BF\ which 
can be associated with
the nodes of the Dynkin diagram $A_{p-1}$. The excitations are the kinks
interpolating between these vacua, and possibly
some neutral particles interpreted as
bound states of the kinks. According to\ \vcomm\ the operators
\vdef\ of the sine-Gordon model are 
related in a simple way to the primary
fields $\Phi_{1,k}$ of\ \mmp
\eqn\relj{V_{1,k}(x) = N_{1,k} \ \Phi_{1,k}(x)\ ,}
where $N_{1,k}$ are numerical factors which depend on the
normalization of $\Phi_{1,k}$. The canonical normalization
\eqn\jshdgt{\langle\,  \Phi_{1,k}(x)\Phi_{1,k}(x')\, \rangle \to
|x-x'|^{-4\Delta_{1,k}} \qquad {\rm as} \qquad |x-x'|\to 0}
corresponds to the choice
\eqn\ksjdgt{
N_{1,k}^2 =
{\cal R}\big(\, {1-k\over 2}\,\beta\,  \big)/ {\cal R}(0)\ ,}
where
$$
{\cal R}(\alpha)=-
\bigg(-{{\pi\mu\, \Gamma(-\beta^2)}\over
{\Gamma(1+\beta^2)}}\bigg)^{1-{1\over \beta^2}-
{2\alpha\over\beta}}\
{{\Gamma\big(\beta^{-2}+2\alpha\beta^{-1}\big)\,
\Gamma\big(\beta^2-2\alpha\beta\big)}\over{\Gamma\big(
2-\beta^{-2}-2\alpha\beta^{-1}\big)\,
\Gamma\big(2-\beta^2+2\alpha\beta\big)}}\ . 
$$
is obtained from\ \sliouville\ by the substitution
\eqn\wystr{b\to i\beta\, ,\ \  a\to i\alpha\, ,\  \
\mu\to-\mu\ .}
Notice that the relation\ \ksjdtr\ can be written as $\lambda=
-N_{1,3}\, \mu$.
With the  normalization\ \jshdgt, one finds
\eqn\evlk{\langle 0_s \mid  \Phi_{1,k} \mid 0_s \rangle =
(-1)^{s(k-1)}\ N_{1,k}
\ {\cal G}\big(\, {1-k\over 2}\, \beta\, \big)\ ,}
where ${\cal G}(\alpha)$ is related
to $G(a)$ in\ \shgev\ by the same substitution\ \wystr.
The sign factor in
\ \evlk\ takes into account the fact that the exponential fields\ \vdef\
with even $k$ change sign when $\varphi$ is translated by the period of
the potential term in\ \sg. 

For the primary fields $\Phi_{l,k}$ with $l>1$ of the restricted 
theory\  \mmp\ the situation is more difficult. The exponential fields
$$
\exp\bigg\{i\Big(  {{l-1}\over 2\beta}-{{k-1}\over 2}\beta\Big)\,
\varphi\bigg\}
$$
for $l>1$ are not invariant with respect to the algebra $U_q
(sl_2)_{+}$. Together with certain nonlocal fields they form
finite-dimensional representations of this algebra. The calculations
become much more involved and we did not complete them yet. However, we
have a conjecture
\eqn\evllk{\langle 0_s \mid  \Phi_{l,k} \mid 0_s \rangle =
{\sin\big(\, {\pi s\over p} |p' l-p k|\, \big)\over
\sin\big(\, {\pi s\over p  } (p'-p)\, \big)}
\ \,  
\biggl[\, M\ {{\sqrt{\pi}\, \Gamma\big(
{3\over 2}+{\xi\over 2}\big)}\over{2\
\Gamma\big({\xi\over 2}\big)}}
\, \biggr]^{2\Delta_{l,k}}\ 
{\cal Q}_{1,3}\big(\, (\xi+1)l-\xi k\, \big)\ ,}
where\ \Zarn
\eqn\sjusy{
M={2\, \Gamma\big({\xi\over 2}\big)
\over \sqrt{\pi}\,
\Gamma\big({1\over 2}+{\xi\over 2}\big) }\
\biggl[\, {\pi\, \lambda\, (1-\xi)(2\xi-1)\over (1+\xi)^2 }\
\sqrt{\Gamma\big({1\over 1+\xi}\big)\,
\Gamma\big({1-2\xi\over 1+\xi}\big)\over
\Gamma\big({\xi\over 1+\xi}\big)\, \Gamma\big({3\xi\over 1+\xi}\big) } 
\  \biggr]^{{1+\xi\over 4}}   }
is the kink  mass and
\eqn\hdyre{\xi={p\over p'-p}\ .}
The function\
${\cal Q}_{1,3}(\eta)$ for $\big|\,\Re e\ \eta\, \big|<\xi$ in\ \evllk\ 
is given by the
integral 
$${\cal Q}_{1,3}(\eta)={\rm exp}\biggl\{
\int_{0}^{\infty} {d t\over t}\ \bigg(\,
{  {\rm cosh}(2 t)\  {\rm sinh}\big( t (\eta-1)\big)\ 
{\rm sinh}\big( t (\eta+1)\big)
\over 2\,  {\rm cosh}(t)\,
{\rm sinh}( t \xi )\,   {\rm sinh}\big(t(1+\xi)
\big)}-
{ (\eta^2-1 )\over 2  \xi (\xi+1)}\,  e^{-4t}\, \bigg)\biggr\} $$ 
and it is defined by
analytic continuation outside this domain
\foot{The expression for $\langle\, \Phi_{l,k}\, \rangle$ proposed 
in\ \LZ\ does 
not contain the first factor\ \evllk\ which carries the dependence 
on $s$. However, the formula in\ \LZ\ is equivalent to\ \evllk\ if the
expectation values in\ \LZ\ are understood not as the matrix elements
between the above ground states $\mid 0_s \rangle$, but rather as the
matrix elements between certain superpositions of these states which
arise in the limit $L\to \infty$ from the asymptotically degenerate
states of the finite-size system, with the spatial coordinate
compactified on a circle of circumference $L$. We will explain this
point elsewhere.}.
In writing\ \evllk\ we
assumed the same canonical normalization convention for
the fields\ $\Phi_{l,k}$\ as in \jshdgt, i.e.
\eqn\ksjdy{\langle 0_s \mid\,
\Phi_{l,k}(x)\Phi_{l,k}(x')\, \mid 0_s \rangle \to
|x-x'|^{-4\Delta_{l,k}} \qquad {\rm as} \qquad |x-x'|\to 0\ .}
Notice that\ \evllk\ automatically satisfy the
relation
$$\langle 0_s \mid  \Phi_{l,k} \mid 0_s \rangle = \langle 0_s
\mid  \Phi_{p-l,p'-k} \mid 0_s \rangle\ .$$ 

\vskip 0.2in

\newsec{Vacuum expectation values  in the Bullough-Dodd model}

\vskip 0.1in

The Bullough-Dodd model is defined by the action\ \Dod, \Zhib
\eqn\bd{{\cal A}_{BD}=\int d^2 x
\Big\{\, {1\over{16\pi}}(\partial_{\nu}\varphi)^2 + \mu e^{b\varphi} +
\mu' e^{-{b\over 2}\varphi}\, \Big\}\ .} 
There is some redundancy in having
two parameters $\mu$ and $\mu'$ in\ \bd\ because if 
one shifts the field variable in\ \bd,
\eqn\shift{\varphi\to\varphi + \varphi_0\ ,}
they change as $\mu \to \mu e^{b\varphi_0}\, ,\
\mu' \to \mu' e^{-{b\over 2}
\varphi_0}$, so that only the combination $\mu\, (\mu')^2$ is invariant.
Nonetheless, we will keep both parameters. In fact in what follows the
combination 
\eqn\m{m={2\, \sqrt3 \ \Gamma\big({1\over 3}\big)\over 
\Gamma\big(
1+{b^2\over 6+3 b^2}\big)\, \Gamma\big({2\over 6+3 b^2}\big)}\ 
\biggl[-{\mu\pi\, \Gamma\big(1+b^2\big)\over
\Gamma\big(-b^2\big)}\, \biggr]^{{1\over  6+3 b^2}}\ 
\biggl[- {2{\mu'}\pi\, 
\Gamma\big(1+{ b^2\over 4}\big)\over
\Gamma\big(-{ b^2\over 4}\big)}\, \biggr]^{{2\over  6+3 b^2}}  }
is proven to be useful. Note that $m$ is invariant under
the shift\ \shift.
The model\ \bd\ is integrable and its factorizable
S-matrix is described in\ \Arien.
It contains a single neutral particle. We
will show below that the mass of this particle  coincides with the
parameter  $m$ defined in\ \m.
In this and the subsequent 
sections we use the notation
\eqn\evbd{G_{BD}(a) = \langle\,  e^{a\varphi}\, \rangle_{BD} }
for the expectation value 
in the Bullough-Dodd model, 
where the exponential field is assumed to be
normalized in accordance with the following
short distance operator product expansion,
\eqn\hsydr{ e^{a\varphi}(x)\, e^{a'\varphi}(x')\to
|x-x'|^{-4 a a'}\ e^{(a+a')\varphi}(x')\ \ \ \ 
{\rm as}\ \ \ \ \ |x-x'|\to 0\ .}
Notice that $|a|$ and $|a'|$ should be sufficiently small numbers, in 
order for\ \hsydr\ to be a leading asymptotic.

Exactly as in the case of the sinh-Gordon model in Sect.2, the Bullough-Dodd
model can be interpreted as the perturbed Liouville theory in  two different
ways, with either $e^{b\varphi}$ or $e^{-{b\over 2}\varphi}$ taken as
the perturbing operator. Using these interpretations and repeating the
arguments in Sect.2 which led to\ \shgra,\ \shgrb,\ one
arrives at two reflection 
relations for\ \evbd
\eqn\bdra{\eqalign{&G_{BD}(a)=R(a)\, G_{BD}(Q-a)\, ,\cr
&G_{BD}(-a)=R'(a)\,  G_{BD}(-Q'+a)\, }}
with
\eqn\QQ{Q={1\over b}+b\, , \qquad Q' = {2\over b}+{b\over 2}\ . }
The function $R(a)$ is exactly the same as in\ \sliouville, while
$R'(a)$ is obtained from that by the substitution $\mu\to\mu'$, 
$b\to b/2$. As in Sect.2, let us now assume that $G_{BD}(a)$ is a
meromorphic function of $a$. Then the following minimal solution to the
equations\ \bdra\ is immediately obtained
\eqn\bdexp{\eqalign{\langle\, e^{a\varphi}\, \rangle_{BD} =& 
\biggl[\, {\mu'\over\mu}\ 
{2^{{b^2\over 2}}\,  \Gamma(1-b^2)\, \Gamma\big(1+
{b^2\over 4}\big)\over  \Gamma(1+b^2)\, \Gamma\big(1-
{b^2\over 4}\big)}\, \biggr]^{{2 a \over 3 b}}\ 
\biggl[{ m\, \Gamma\big(
1+{ b^2\over 6+3 b^2}\big)\, \Gamma\big({2\over 6+3 b^2}\big)\over
2^{{2\over 3}}\, 
\sqrt3\   \Gamma\big({1\over 3}\big)}\biggr]^{a b-2 a^2}
\times\cr
&\exp\biggl\{\int_0^{+\infty}\, {d t\over t}\ \Big(
-{{\sinh\big( (2+b^2) t\big)\, 
\Psi(t,a)}\over {\sinh\big(3 (2+b^2) t\big) \, \sinh(2t) \, 
\sinh\big(b^2 t\big)}}+2 a^2\, e^{-2t}\, \Big)\biggr\} \ ,}}
where
$$
\eqalign{\Psi(t,a)=&   \sinh\big(2a b t\big)\,
\Big(\,
\sinh\big( (4+ b^2+2a b) t\big)-
\sinh\big( (2+2b^2-2a b) t\big)+\cr &
\sinh\big( (2+ b^2+2a b) t\big)-
\sinh\big( (2+ b^2-2a b) t\big)-
\sinh\big( (2- b^2+2a b) t\big)\, \Big)\ .}
$$
and $m$ is given by\ \m.
The integral in\ 
\bdexp\ is convergent if 
$$ -{1\over b}-{b\over 4}<\Re e\,  a <{1\over 2 b}+{b\over 2} \ ;$$
it should be
understood in terms of analytic continuation otherwise. We propose
\bdexp\ as exact expectation values for the Bullough-Dodd model\ \bd.

Expanding\ $\langle\,  
e^{a\varphi}\,  \rangle_{BD} = 1 + a\,  \langle\,
\varphi\,  \rangle_{BD} + O(a^2)$,
one finds
the expectation value of
$\varphi$,
\eqn\phiexp{\eqalign{&
\langle\,\varphi\, \rangle_{BD}={2\over 3 b}\ \log\biggl\{\, 
{\mu'\over \mu}\
{\Gamma(1-b^2)\, \Gamma\big(1+
{b^2\over 4}\big)\over  \Gamma(1+b^2)\, \Gamma\big(1-
{b^2\over 4}\big)}\ 
\bigg[ { m\, \Gamma\big(
1+{ b^2\over 6+3 b^2}\big)\, \Gamma\big({2\over 6+3 b^2}\big)\over
2^{{1\over 3}}\,
\sqrt3\   \Gamma\big({1\over 3}\big)}\bigg]^{{3 \over 2 } b^2}\, \biggr\}  
-\cr
&8 b\ \int_{0}^{+\infty} d t\, 
{\sinh\big((2+b^2) t\big)\, \sinh\big((1-{b^2\over 2})t\big)\, 
\sinh\big((2+{b^2\over 2})t\big)
\, \sinh\big((1+b^2)t\big)
\over \sinh\big(3 (2+b^2) t\big) \, \sinh(2t) \,
\sinh\big(b^2 t\big)}\ .}}
For $a=b$ and for $a=-b/2$ the integral in\ \bdexp\ can be
evaluated explicitly,
\eqn\hsydr{\mu\, \langle\ e^{b\varphi}\, \rangle_{BD}=
 \mu'/2\ \langle\, e^{- {b\over 2}\varphi}\, \rangle_{BD}= 
{m^{2}\over  24\, \sqrt3\,  (2+b^2)\, 
\sin\big(
{\pi b^2\over 6+3 b^2}\big) \, \sin\big(
{2\pi \over 6+3 b^2}\big)}\ .}
These expectation values can be used to derive the bulk specific free
energy of the Bullough-Dodd model
\eqn\shydt{f_{BD} =- \lim_{V\to\infty} {1\over V}\, \log Z_{BD}\ ,}
where $V$ is the volume of the 2D space and $Z_{BD}$ is the
singular part of the partition function associated with\ \bd.
Obviously,
\eqn\jsydtr{
\partial_{\mu} f_{BD}=\langle\, e^{b\varphi}\,
\rangle_{BD}\, ;\qquad
\partial_{\mu'} f_{BD}=
\langle\, e^{-{b\over 2}\varphi}\,\rangle_{BD}\ .}
This leads to the following result,
\eqn\freeen{f_{BD}={m^2\over 16\sqrt3\, \sin\big(
{\pi b^2\over 6+3 b^2}\big) \, \sin\big(
{2\pi \over 6+3 b^2}\big)}\ .}
On the other hand exact expression for  the specific free energy
in terms of the physical particle mass can
be obtained from exact form-factors\ \acerbi,\ \brl, or from
the Thermodynamic Bethe Ansatz calculations following\ \Zar,\ \fatt.
This way one obtains exactly\ \freeen\ with $m$ understood as the
particle mass. This shows that\ \m\ indeed gives the particle mass
in the Bullough-Dodd model.

Since the above derivation of\ \bdexp\ is based on the assumptions 
(in particular, we made strong analyticity assumption), some checks of 
this result are desirable. Simple consistency check is based on the
known fact that for $b^2=2$ the Bullough-Dodd model is equivalent to the
sinh-Gordon model\ \shg\ with $b^2 = 1/2$. It is possible to check that
\ \bdexp\ calculated at $b^2=2$ coincides with\ \shgev\ 
calculated at $b^2
= 1/2$.

Important check can be performed in the classical limit $b^2\to 0$.
Consider the expectation value $\langle\, e^{{\sigma\over
b}\varphi}\, \rangle_{BD}$. In the limit $b^2\to 0$ with $\sigma$ fixed,
\bdexp\ gives
\eqn\bdexpcl{\log\,  \langle\,  e^{{\sigma\over b}\varphi}\, 
\rangle_{BD} = {2\over
b^2}\biggl(- \sigma^2\,  \log m + {\sigma\over 3}\, \log
\Big(\,  {\mu'\over2\mu}\, \Big) +
\int_{0}^{\sigma} d\omega \, C(\omega)\biggr)+ O(1)\ ,}
where
\eqn\Cbd{C(\omega)=\big(\,  2\log 2+3\log 3\, \big)\, \omega+ \log\biggl[\,
{\Gamma\big({1+\omega\over 3}\big)\, \Gamma\big({2+2\omega\over 3}\big)
\over \Gamma\big({2-\omega\over 3}\big)\,
\Gamma\big({1-2\omega\over 3}\big)}\, \biggr]\ .}
On the other hand, for $b^2\to 0$ the expectation value\ 
$\langle\,  e^{{\sigma\over b}\varphi}(0)\, \rangle_{BD}$\ can
be calculated
directly in terms of the action\ \bd\ evaluated 
on appropriate classical solution $\varphi_{cl}(x)$ of the equations of 
motion associated with\ \bd. The suitable solution depends only on the
radial coordinate $r=|x|$. It can be written as\ 
$\varphi_{cl}(r) = {2\over  b }\, \big(  \phi(r)+
{1\over 3} \log( {\mu'\over 2\mu} )\, \big)$,\ where $\phi(r)$ is the
solution to the classical Bullough-Dodd equation
\eqn\bdclass{\partial_r^2\phi+r^{-1}\partial_r\phi={m^2\over 3}\, 
\Big(\, e^{2\phi}-e^{-\phi}\, \Big)\  ,} 
which satisfies the following asymptotic conditions
\eqn\bdas{\eqalign{&
\phi(r)\to -2\sigma\, \log (m r)+{\tilde C}(\sigma)\ \ \ \ 
{\rm as}\ \ r\to 0\ ,\cr &
\phi(r)\to {4 \sqrt3 \over \pi}\ \sin\big({\pi\sigma\over 3}\big)\, 
\sin\big({\pi\over 3} (1-\sigma)\big)\  K_0(mr) \ \ \ \
{\rm as}\ \ r\to +\infty\ ,}}
where $K_0(t)$ is the MacDonald function.
The constant term ${\tilde C}(\sigma)$ in\ \bdas\ is not arbitrary but
must be consistently determined from the equation\ \bdclass. Exact
result for ${\tilde C}(\sigma)$ found in\ \kitaev\ 
(see also \tracy)\ shows
that it
coincides with $C(\sigma)$ defined by\ \Cbd. Then
calculation of the classical action gives the result identical to
\bdexpcl. 

One can go beyond the classical limit and consider the loop expansion
for the expectation values\ \evbd. The simplest thing to study is the
expectation value $\langle\,  \varphi\, \rangle_{BD}$.
According to\ \phiexp,
this quantity can be written as a power series in $b$,
\eqn\phiser{\langle\,  \varphi\,  \rangle_{BD} =
{2\over 3 b}\, \log\Big(\, {\mu'\over 2\mu}\, \Big)+
b\, \big(\, \gamma+\log(m/ 2)\, \big)+{b^3\over 108}\,
\big(\, 3\sqrt3\, \pi+10\pi^2-15\,\psi'\big(2/3\big)\, \big)+
O(b^5)\, , }
where\ $\psi'(t)=\partial^2_t\log\big(\Gamma(t)\big)$\ and\ 
$\gamma=0.577216...$\ is Euler's constant.
On the other hand, it is possible to calculate this expectation value
within the standard Feynman perturbation theory for the action\ \bd. 
The first term in\ \phiser\ coincides with the classical value of
$\varphi$, and the diagrams contributing to the next two orders are
shown in Fig.1. 
Calculations are straightforward and the result is in 
exact agreement with\ \phiser. 
\midinsert

\centerline{\epsfbox{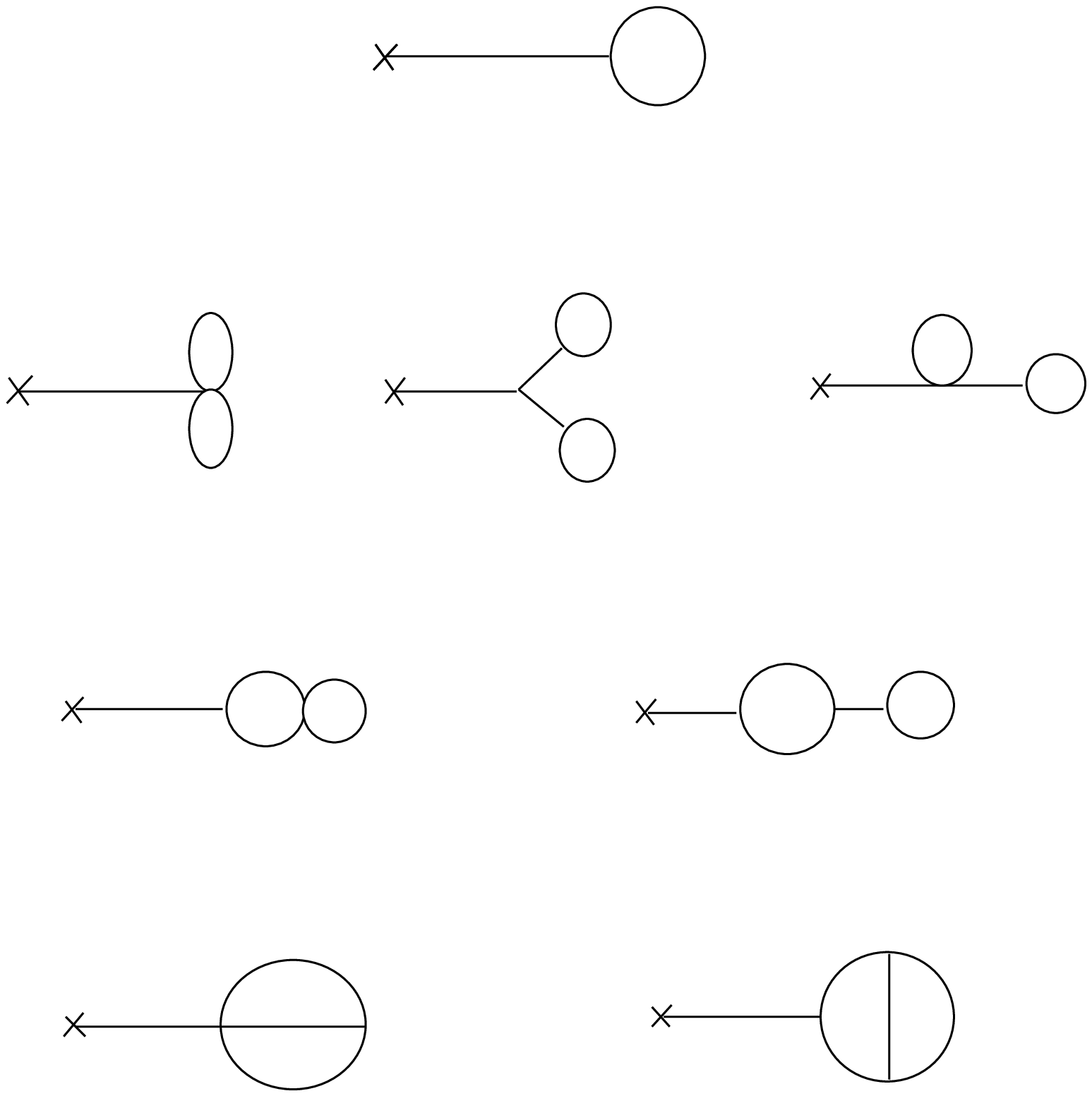}}
\vskip 7 truemm
\capt{
Fig.1. Feynman diagrams contributing 
to $ \langle\,  \varphi\,  \rangle_{BD}$ to the orders $b$ and $b^3$.}
\vskip 7 truemm
\endinsert

\vskip 0.2in

\newsec{Expectation values of primary fields 
in the minimal models perturbed by
the operator $\Phi_{1,2}$}

\vskip 0.1in

The expectation value\ \bdexp\ proposed in the previous section allows
one to obtain the the expectation values 
\eqn\cbdev{{\cal G}_{cBD}(\alpha) = \langle\, 
e^{i\alpha\varphi}\, \rangle_{cBD} }
in the so called ``complex Bullough-Dodd model'',
\eqn\cbd{{\cal A}_{cBD} = \int d^2 x
\Big\{\, {1\over{16\pi}}(\partial\varphi)^2 - \mu e^{i\beta\varphi}- 
\mu' e^{-i{\beta\over 2}\varphi}\, \Big\}\ ,} 
which is obtained from\ \bd\ by replacing the parameter $b$ by pure
imaginary value $b\to i\beta$ and $\mu\to-\mu, \ 
\mu'\to-\mu'$. The action\ \cbd\ is complex and it is not
clear exactly how it defines a quantum field theory. Nonetheless, some
formal manipulations can be done. In particular, 
the model\ \cbd\ is shown
to be integrable and its factorizable 
S-matrix is identified in\ \smir.
Although the model\ \cbd\ is very
different from\ \bd\ in its physical content 
(the model\ \cbd\ contains
solitons), there are good reasons to believe that
the expectation values\ \cbdev\ are obtained from the
expectation values\ \bdexp\ by simple substitution 
\eqn\gg{{\cal G}_{cBD}(\alpha) = G_{BD}(i\alpha)\big|_{b=i\beta}\ .}
The arguments are
much the same as those which lead us to the relation
between the expectation values in\ \shg\ and\ \sg. Namely, it is easy to
check that for fixed $a$ the expression\ \bdexp\ can be expanded into a
power series in $b$ with a finite radius of convergence. This suggests
that in principle\ \bdexp\ can be calculated by summing up the Feynman
perturbation theory series for\ \bd. At the same time the perturbation
theory for\ \bd\ agrees with that for\ \cbd\ to all orders if one makes
the substitution $b \to i\beta$. 

The complex Bullough-Dodd model admits the quantum group restriction 
similar to the one in the sine-Gordon model\ \smir. 
There are some differences,
though. The model\ \cbd\ (in the infinite space) has
a symmetry with respect to
the affine quantum group algebra $U_q (A_2 ^{(2)})$ where $q$ is given
by the same expression\ \qdef. This algebra contains a subalgebra 
$U_q (sl_2)$ which can
be used for the quantum group restriction of\ \cbd
\foot{The
above affine quantum algebra contains another subalgebra,
$U_{q^4}(sl_2)$, which can be used for another quantum group
restriction of\ \cbd. We  briefly discuss this case in Sect.5.}. If 
$\beta^2$ takes the rational values\ \betap\ the restricted
theory coincides with perturbed minimal model\ \smir
\eqn\mp{{\cal M}_{p/p'}+\lambda \int d^2 x \,\Phi_{1,2}(x)\ .}
with 
$$
\lambda^2=-{\pi\ \mu\, (\mu')^2\over (2\beta^2-1)^2}\ 
{\Gamma^2(1-\beta^2)\,\Gamma(2\beta^2)
\over\Gamma^2(\beta^2)\, \Gamma(1-2\beta^2)}\ .
$$
As in\ \sg, the generators $E_{+}, H_{+}, F_{+}$ of the subalgebra $U_q
(sl_2)$ commute with the exponential fields\ \vdef\ which become the
primary fields  $\Phi_{1,k}\ ( k=1,2,...,p'-1)$ in the restricted 
theory\ \mp. Therefore, the same arguments as in Sect.2 lead to the
following expression for the expectation values of these primary fields 
in\ \mp,
\eqn\bdlk{\langle 0_s \mid \Phi_{1,k}\mid 0_s \rangle =
(-1)^{s(k-1)}\ N_{1,k}\ {\cal G}_{cBD}\big(\, 
{1-k\over 2}\, \beta\, \big)\ , }
which is similar to\ \evlk,  except
that now ${\cal G}_{cBD}(\alpha)$ stands
for the
expectation value\ \cbdev\ in the complex Bullough-Dodd model. Here again,
the canonical normalization\ \ksjdy\ of the primary fields $\Phi_{l,k}$ is
assumed. In\ \bdlk\  $s$ is an integer which labels the ground states of 
the perturbed
theory. Precisely which values it takes relates to the question of
the vacuum structure of the perturbed theory\ \mp; we will discuss this
point a little later. For now,
it suffices to note that only the signs of
the the expectation values\ \bdlk\ depend on the choice of the vacuum. 

Particular case
of\ \bdlk\ is the expectation value of the perturbing operator,
\eqn\mplll{\eqalign{\langle 0_s \mid \Phi_{1,2}\mid 0_s \rangle =
(-1)^{s-1}&\ |\lambda|^{{\xi-2\over 3\xi+6}}\ \, 
{2^{{2\xi+10\over 3\xi+6}}\, (\xi+1)\,  
\Gamma^2\big({1\over 3}\big)\over \sqrt3\, (\xi+2)\, \pi^2}\ 
{\Gamma\big({\xi+4\over 3\xi+6}\big)\, 
\Gamma\big({\xi\over 3\xi+6}\big)\over
\Gamma\big({2\xi+2\over 3\xi+6}\big)\,
\Gamma\big({2\xi+6\over 3\xi+6}\big)}\times\cr & 
\biggl[\,  {\pi^2\,  \Gamma^2\big({3\xi+4\over 4\xi+4}\big)\,
\Gamma\big({1\over 2}+{1\over \xi+1}\big)\over
\Gamma^2\big({\xi\over 4\xi+4}\big)\,
\Gamma\big({1\over 2}-
{1\over \xi+1}\big) }\, \biggr]^{{2\xi+2\over 3\xi+6}}\ ,}}
where $\xi$ is given by\ \hdyre.
As usual, this expectation value is related to the specific free energy
$f_{1,2}$ of the perturbed theory\ \mp 
\eqn\flambda{\partial_{\lambda}
f_{1,2}= \langle 0_s \mid
\Phi_{1,2}\mid 0_s \rangle\ .}
On the other hand, the specific free energy for\ \mp\ is known exactly in
terms of the mass $M$ of the lightest kink present in this theory\ \fatt,
\eqn\fM{f_{1,2} =-{M^2\  \sin\big({\pi\, \xi\over 3\xi+6}\big)
\over 4\, \sqrt3\,
\sin\big({\pi (2\xi+2)\over 3\xi+6}\big)}\  . }
Combining\ \mplll,\ \flambda\ and\ \fM, one finds the relation between the
coupling constant $\lambda$ in \mp\ and the mass $M$,
\eqn\Mlambda{M={2^{{\xi+5\over 3\xi+6}}\, \sqrt3\, 
\Gamma\big({1\over 3}\big)\, 
\Gamma\big({\xi\over 3\xi+6}\big)
\over\pi\, 
\, \Gamma\big({2\xi+2\over 3\xi+6}\big)}\
\biggl[\,  {\pi^2\, \lambda^2\, \Gamma^2\big({3\xi+4\over 4\xi+4}\big)\,
\Gamma\big({1\over 2}+{1\over \xi+1}\big)\over
\Gamma^2\big({\xi\over 4\xi+4}\big)\,
\Gamma\big(
{1\over 2}-{1\over \xi+1}\big) }\, \biggr]^{{\xi+1\over 3\xi+6}}\ ,}
in exact agreement with\ \fatt. According to\ \Mlambda,
the perturbed QFT\ \mp\ develops a  massive spectrum for 
\eqn\jsudt{\eqalign{&0<\xi<1\, \ \ \ \ \ \ \ \Re e\, \lambda=0\, ;\cr
&\xi>1\, ,\ \ \ \ \ \ \ \ \ \ \ \Im m\, \lambda=0\, .}}
In what follows we will discuss the second case only.

As in the case of $\Phi_{1,3}$ perturbation the situation with other
primary fields $\Phi_{l,k}$ with $l>1$ in\ \mp\ is more difficult.
However, there is a natural modification of the conjecture\ \evlk\
suitable for\ \mp,
\eqn\bdlk{
\langle 0_s \mid  \Phi_{l,k} \mid 0_s \rangle =
{\sin\big(\, {\pi s\over p} |p' l-p k|\, \big)\over
\sin\big(\, {\pi s \over p  } (p'-p)\, \big)}
\ \,
\biggl[\,  {M\, \pi\, (\xi+1)\, \Gamma\big({2\xi+2\over 3\xi+6}\big)\over
2^{{2\over 3}}\, \sqrt3\, \Gamma\big({1\over3}\big)\,
\Gamma\big({\xi\over 3\xi+6}\big)}
\, \biggr]^{2\Delta_{l,k}}\ {\cal Q}_{1,2}\big(\, (\xi+1)l-\xi k\, \big)\ ,}
where ${\cal Q}_{1,2}(\eta)$ for $\big|\,\Re e\ \eta\, \big|<\xi\ (\xi>1)$
is given by the
integral
$$\eqalign{{\cal Q}_{1,2}&(\eta)={\rm exp}\biggl\{
\int_{0}^{\infty} {d t\over t}\ \bigg(\,
{  {\rm sinh}\big((\xi+2) t\big)\  {\rm sinh}\big( t (\eta-1)\big)\
{\rm sinh}\big( t (\eta+1)\big)
\over   {\rm sinh}\big(3(\xi+2) t\big)\, {\rm sinh}\big(2 (\xi+1) t\big)\,
{\rm sinh}(  \xi t )
}\times\cr
&\Big( \cosh\big(3(\xi+2) t\big)+
\cosh\big((\xi+4) t\big)-
\cosh\big((3\xi+4) t\big)+\cosh\big(\xi t\big)+1 \Big)
-\cr
&\ \ \ \ \ \ \ \ \ \ \ \ \ \ \ \ \ \ \ \ \ \ \ \ \ \ \ \ \ \ \ \ \
{ (\eta^2-1 )\over 2  \xi (\xi+1)}\,  e^{-4t}\, \bigg)\biggr\}\ ,  }$$
and  is defined
through analytic continuation outside this domain.
The integer $s$ in\ \bdlk\ labels the vacuum states of\ \mp. To
discuss the vacuum structure of\ \mp\ let us recall the basic idea of 
quantum group restriction. The states of\ \cbd\ can be
classified according to the representations of the above quantum algebra
$U_q (sl_2)$. If $\beta^2$ takes the rational value\ \betap\ 
the subspace ${\cal H}_{p}\in {\cal H}_{cBD}$ consisting of the
representations with the spins $j=0, 1/2, 1, ..., p/2-1$, is closed
with respect to the dynamics of\ \cbd. However unlike the sine-Gordon
case, the solitons of\ \cbd\ transform as the three-dimensional
representations of $U_q (sl_2)$ with the spin $j=1$.
Therefore in fact there are two dynamically closed subspaces, ${\cal
H}_{p}^{+}$ and ${\cal H}_{p}^{-}$, containing respectively
half-integer or
integer spins out of the above set $j=0,1/2,...,p/2-1$.
Each of the spaces $Inv\big({\cal H}^{+}_{p}\big)$
and $Inv\big({\cal H}^{-}_{p}\big)$ of  invariant tensors
associated with ${\cal H}_{p}^{+}$ and 
${\cal H}_{p}^{-}$ can be
interpreted as the space of states of certain quantum field
theory. Therefore, this quantum group restriction of\ \bd\ gives rise to
two different quantum field theories. Notice that if $p$ is odd, the spaces
$Inv\big({\cal H}^{+}_{p}\big)$\
and $Inv\big({\cal H}^{-}_p\big)$ are isomorphic because of the
known property  of the  tensor category of  representations of
$U_q(sl_2)$ with $q^{p}=\pm 1$,
and hence the
corresponding field theories are equivalent. If $p$ is even, these are
two really different field theories. Let us recall in this connection
that the minimal CFT ${\cal M}_{p/p'}$ always has $Z_2$ symmetry which 
acts on the primary fields as
\eqn\ztwo{\Phi_{l,k} \to (-1)^{(l-1)p'-(k-1)p}\ \Phi_{l,k}\  .}
If $p$ is odd, the perturbing operator $\Phi_{1,2}$ is odd under the
transformation\ \ztwo, and 
therefore changing the sign of the perturbation
in\ \mp\ leads to an equivalent field theory -- all its correlation
functions are obtained from those of the original theory by the
substitution\ \ztwo. On the contrast, if $p$ is even, the operator
$\Phi_{1,2}$ is invariant under\ \ztwo; in this case
the theories\ \mp\ with
different signs of the perturbation are expected to be essentially
different\foot{The exception is the case $(p,p')=(4,5)$ where
the theories with different signs of $\lambda$ in\ \mp\ are related by
duality transformation\ \sas,\ \carmus.}. 
Therefore it is natural to identify 
$Inv\big({\cal H}_{p}^+\big)$ with the 
space of states of\ \mp\ with $\lambda > 0$ and $
Inv\big({\cal H}^{-}_{p}\big)$ with the 
space of states of\ \mp\ with $\lambda < 0$. The integers $s$ labeling
the vacua in\ \bdlk\ are related to the spins $j$ admitted into ${\cal
H}_{p}$ as $s=2j+1$. We conclude that the theory\ \mp\ with odd $p$
has ${p-1}\over 2$ degenerate ground states independently on the sign of
$\lambda$; however, if $\lambda > 0$ these ground states are identified
with the above vacua $\mid 0_s \rangle$ with 
even  $s=2,4,...,p-1$, while if 
$\lambda < 0$ these are the vacua $\mid 0_s \rangle$ with  odd
$s=1,3,...,p-2$. If $p$ is even and $\lambda > 0$ there are $p/2-1$
ground states identified with $\mid 0_s \rangle$ with even
$s=2,4,...,p-2$. Finally, if $p$ is even and $\lambda < 0$ the
theory\ \mp\ has $p/2$ ground states $\mid 0_s \rangle$ with even
$s=1,3,...,p-1$. With this understanding, the formula\ \bdlk\ applies to
all models\ \mp, both with positive and negative $\lambda$.

In\ \magnol\ the Truncated Conformal Space method\ \Yur\ was
adopted to obtain
numerically the expectation values of primary fields in the perturbed
theory\ \mp\ for some $(p,p')$. It is interesting to compare our results
to these numerical data.

The model\ \mp\ with $(p,p')=(3,4)$ describes the Ising
model at critical temperature with nonzero magnetic field. In this case
there is only one vacuum $|0\rangle \equiv |0_2 \rangle$ (we assume that
$\lambda > 0$), and\ \bdlk\ gives
$$
\eqalign{&\langle 0 \mid\,\Phi_{1,2}\,\mid 0 \rangle=-1.27758...
\ \lambda^{1\over 15}\ ,\cr
&\langle 0 \mid\,\Phi_{1,3}\,\mid 0 \rangle=
2.00314...\  \lambda^{8\over 15}\  .}
$$
According to numerical calculations in this case
$$
\eqalign{&\langle 0 \mid\, \Phi_{1,2}\, \mid 0 \rangle_{num}=
-1.277(2)\ \lambda^{1\over 15}\ ,\cr
&\langle 0 \mid\, \Phi_{1,3}\, \mid 0 \rangle_{num}=
1.94(6)\  \lambda^{8\over 15}\ . }
$$
The VEV $\langle 0 \mid\Phi_{1,3}\mid 0 \rangle$ was also obtained 
from  the fit of lattice data with the
result $2.02(10)\ \lambda^{8\over 15}$\ \magnol.

The case $(p,p')=(4,5)$ in\ \mp\ describes Tricritical Ising model
perturbed by the leading energy density operator $\varepsilon(x)$ of the
conformal dimension $\Delta_{1,2} ={1\over 10}$. The minimal model
${\cal M}_{4/5}$ contains four more primary fields (besides the above
field $\varepsilon = \Phi_{1,2}$ and the identity operator), the
sub-leading energy density operator $\varepsilon' = \Phi_{1,3}$
with $\Delta_{1,3}={3\over 5}$ (sometimes referred to as the 
``vacancy operator''), two magnetic operators
$\sigma=\Phi_{2,2}\  (\Delta_{2,2}={3\over 80}$) and $\sigma' =
\Phi_{2,1}\   (\Delta_{2,1}={7\over 16}$) and $\Phi_{1,4}\ $ 
(the
latter does not have obvious physical interpretation). If $\lambda < 0$
the Ising symmetry $\sigma \to -\sigma$ is spontaneously broken and
there are two ground states $\mid + \rangle \equiv \mid 0_1 \rangle$ and
$\mid - \rangle \equiv \mid 0_3 \rangle$.
If $\lambda>0$ there is a  unique
ground state $\mid 0 \rangle \equiv \mid 0_2 \rangle$.
With these values of $s$
\bdlk\ gives
$$
\eqalign{&\langle \pm \mid \,\Phi_{1,2}\,\mid \pm \rangle=
 1.46840...\ (-\lambda)^{{1\over 9}} \ \ \ \ \ \ \ \ \  
\langle 0 \mid \,\Phi_{1,2}\,\mid 0 \rangle =- 1.46840...
\ \lambda^{{1\over 9}}\cr
&\langle \pm \mid \,\Phi_{1,3}\,\mid \pm \rangle =
3.70708...\ (-\lambda)^{{2\over 3}} \ \ \ \ \ \ \ \ \ 
\langle 0 \mid \,\Phi_{1,3}\,\mid 0 \rangle=
3.70708...\ \lambda^{{2\over 3}}\cr
&\langle \pm \mid \, \Phi_{2,2}\,\mid \pm \rangle = 
\pm 1.59427... 
\ (-\lambda)^{{1\over 24}}
 \ \ \ \ \ \ 
\langle 0 \mid \,\Phi_{2,2}\,\mid 0 \rangle=0\ \cr
&\langle \pm \mid \, \Phi_{2,1}\,\mid \pm \rangle = 
\pm 2.45205...
\ (-\lambda)^{{35\over 72}}
\ \ \ \ \ \ 
 \langle 0 \mid \,\Phi_{2,1}\,\mid 0 \rangle=0\ .}
$$
These numbers can be compared 
with the numerical results quoted in\ \magnol
$$
\eqalign{&\langle \pm \mid \,\Phi_{1,2}\,\mid \pm \rangle_{num}=
1.466(6)\ (-\lambda)^{1\over 9}\,\ \ \ \ \ \ \ 
\langle 0 \mid \,\Phi_{1,2}\,\mid 0 \rangle_{num} =- 1.465(5)
\ \lambda^{{1\over 9}}\cr
&\langle \pm \mid \,\Phi_{1,3}\,\mid \pm \rangle_{num}=
3.5(3)\ (-\lambda)^{2\over 3}\, \ \  \ \ \ \ \ \ \ \  
\langle 0 \mid \,\Phi_{1,3}\,\mid 0 \rangle_{num}=
3.4(2)\ \lambda^{2\over 3}\cr
&\langle \pm \mid \,\Phi_{2,2}\,\mid \pm \rangle_{num}=
\pm 1.594(2)\ (-\lambda)^{{1\over 24}}\ \ \ \ 
\langle 0 \mid \,\Phi_{2,2}\,\mid 0 \rangle_{num}=0\cr
&\langle \pm \mid \,\Phi_{1,2}\,\mid \pm \rangle_{num}=
\pm 2.38(6)\ (-\lambda )^{{35\over 72}}\ \ \ \ \
\langle 0 \mid \,\Phi_{2,1}\,\mid 0 \rangle_{num}=0\  .}
$$
The VEV $\langle \,\Phi_{1,3}\, \rangle$ was earlier 
estimated in the work\ \mag\ as\
$3.78\ |\lambda|^{2\over 3}$, with the quoted    error of\ $5-10\%$

\vskip 0.2in

\newsec{Minimal models perturbed by the operators $\Phi_{1,5}$}

\vskip 0.1in

The affine symmetry algebra $U_q (A_{2}^{(2)})$ of\ \cbd\ contains also 
the algebra $U_{q^4}(sl_2)$ as a subalgebra. One can use it to obtain 
another quantum
group restriction of\ \cbd. Let $p, p'$ be two relatively prime integers
such that $2p < p'$. As is known, for
\eqn\betapp{\beta^2 = {{4 p}\over p'} }
this restriction gives the perturbed minimal model
\eqn\mpp{{\cal M}_{p/p'} + {\bar \lambda}\int d^2 x\,  \Phi_{1,5} \ .}
The above condition $2p < p'$ (which excludes unitary models ${\cal
M}_{p/p+1}$) guarantees that the perturbation is relevant. The coupling
parameter $\bar \lambda$ in\ \mpp\ is related to the parameters $\mu,
\mu'$ in\ \cbd\ as 
\eqn\kdidy{{\bar \lambda}^2=\biggl[\, 
{32\, \pi^2\ \mu\ (\mu')^2
\over(4-5\beta^2)(1-\beta^2)(4-3\beta^2)(2-\beta^2)}\, \biggr]^2\ 
{\Gamma^5\big(1-{\beta^2\over 4}\big)\, \Gamma\big({5\beta^2\over 4}\big)
\over
\Gamma^5\big({\beta^2\over 4}\big)\, 
\Gamma\big(1-{5\beta^2\over 4}\big)}\ }
According to the
general scheme of the quantum group restriction one expects that the
restricted theory has a particle of the mass $m$
(possibly among other particles and kinks) given by\ \m\ with $b^2$
replaced by $-\beta^2$\ and $\mu\to-\mu,\ 
\mu'\to-\mu'$.
Excluding $\mu \, (\mu')^2$ from these relations,
we can expresses the perturbation parameter ${\bar \lambda}$
in terms of the physical mass scale $m$,
\eqn\hsydr{m={2 \sqrt3\, \Gamma\big(
{1\over 3}\big)\over \Gamma\big({3-5\xi\over 3-3\xi}\big)\,
\Gamma\big({1+\xi\over 3-3\xi}\big)}\
\biggl[{ 4 \pi^2\, {\bar \lambda}^2\, (1-4\xi)^2 (1-2\xi)^2 \,
\Gamma^2\big({3-\xi\over 1+\xi}\big)\, \Gamma\big({\xi\over 1+\xi}\big)\,
\Gamma\big({1-4\xi\over 1+\xi}\big)\over
(1+\xi)^4\ \Gamma^2\big({4\xi\over 1+\xi}\big)\,
\Gamma\big({1\over 1+\xi}\big)\,
\Gamma\big({5\xi\over 1+\xi}\big)}\, \biggr]^{{1+\xi\over 12(1-\xi)}}\ .}
Here  we use the notation
$$\xi = {p\over {p'-p}}\ . $$ 
As it follows from\ \hsydr, QFT\  
\mpp\ presumably  has a massive spectrum  for 
\eqn\jshdt{\eqalign{&0 < \xi < {1\over 4}\, ,\ \ \ \ \ 
\Im m\,   {\bar\lambda}=0\ ,\cr
&{1\over 4} < \xi < {3\over 5}\, ,\ \ \ \ \
\Re e\,  {\bar\lambda}=0\ .} }
Outside this domain the physical content of the model\ \mpp\
is particularly unclear. We restrict our following
discussion to the domain\ \jshdt.
Then, the specific free energy of\ \mpp\ can\ be
obtained from\ \freeen\ by
the  substitution $b^2\to- {4\xi\over 1+\xi}\, $, i.e.
\eqn\freep{f_{1,5}=-{m^2\over 16\, \sqrt3\,
\sin\big({2\pi \xi\over3- 3\xi}\big)
\, \sin\big({\pi (1+\xi)\over 3-3\xi}\big)}\  .}
Using this relation and\ \hsydr\ one
derives  the following expression for the
expectation value of the perturbing operator in\ \mpp,
\eqn\jshdfgr{\eqalign{
\langle\,\Phi_{1,5}\, \rangle =&
{ \pi\, (1-4\xi)(1-2\xi) \
\Gamma\big({2-2\xi\over1+\xi}\big)\over
12\sqrt3\, (1+\xi)^2\, \Gamma\big({4\xi\over1+\xi}\big)\
\sin\big({2\pi\xi\over 3-3\xi}\big)
\sin\big({\pi(1+\xi)\over 3-3\xi}\big)}\
\sqrt{{\Gamma\big({\xi\over 1+\xi}\big) 
\Gamma\big({1-4\xi\over 1+\xi}\big)
\over \Gamma\big({1\over 1+\xi}\big)
\Gamma\big({5\xi\over 1+\xi}\big)}}\times\cr
&\biggl[\,  {    \Gamma\big({3-5\xi\over 3- 3\xi}\big)
\, \Gamma\big({1+\xi\over 3-3\xi}\big)\over 2 \,
\sqrt3\,  \Gamma\big({1\over 3}\big)}
\, \biggr]^{{6\xi-6\over 1+\xi}}\ m^{{8\xi-4\over 1+\xi}}\ .}}

To obtain more general expectations values, let us note that
the generators
$E_{-}, H_{-}, F_{-}$ of $U_{q^4}(sl_2)$ commute with the exponential
fields $e^{i{{k-1}\over 4}\varphi}(x)$, $ k=1,2,...$ which under the
restriction become primary fields $\Phi_{1,k}(x)$ in the perturbed
theory\ \mpp. Using the results of Sect.3 and 4 we can 
obtain
\eqn\philksw{\langle 0_s \mid \Phi_{1,k} \mid 0_s \rangle =
(-1)^{(k-1)s}\ \biggl[\, 
{m\,  (1+\xi)\,  \Gamma\big({3-5\xi\over 3- 3\xi}\big)
\, \Gamma\big({1+\xi\over 3-3\xi}\big)\over 2^{8\over 3}\,
\sqrt3\,  \Gamma\big({1\over 3}\big)}
\, \biggr]^{2\Delta_{1,k}}\ {\cal Q}_{1,5}\big((1+\xi-\xi k\big)\ ,}
where\ ${\cal Q}_{1,5}(\eta)={\cal Q}(\eta)/{\cal Q}(1)$ and
the function\  
${\cal Q}(\eta)$ for $\big|\,\Re e\ \eta\, \big|<\xi\ (\xi<1)$
is given by the
integral
$$\eqalign{{\cal Q}&(\eta)={\rm exp}\biggl\{
\int_{0}^{\infty} {d t\over t}\ \bigg(\,
{  {\rm sinh}\big((1-\xi) t\big)\  {\rm cosh}\big( 2t \eta\big)
\over2\,    {\rm sinh}
\big(3(1-\xi) t\big)\, {\rm sinh}\big( (1+\xi) t\big)\,
{\rm sinh}( 2 \xi  t)
}\times\cr
&\Big( \cosh\big(3(1-\xi) t\big)+
\cosh\big((3-\xi) t\big)-
\cosh\big((1-3\xi) t\big)+\cosh\big((1+\xi) t\big)+1 \Big)
-\cr
&\ \ \ \ \ \ \ \ \ \ \ \ \ \ \ \ \ \ \ \ \ \ \ \ \ \ \ \ \ \ \ \ \
{ (\eta^2-1 )\over 2  \xi (\xi+1)}\,  e^{-4t}\, \bigg)\biggr\}\ .  }$$
In\ \philksw,\ 
again $s$ is an integer labeling the ground states of \mpp. The
ground state structure of\ \mpp\ is not completely understood and we do not
discuss it here. Note that the value of $s$ affects only the sign of the
expectation values\ \philksw\ with even $k$.

Some of the perturbed theories\ \mpp\ are studied in the literature\ 
\mus,\ \martins,\
\takacs. Let us see how our result\ \hsydr\ matches the data from these
references.

1. The model\ \mpp\ with $(p,p')=(2,7)$ was studied in\ \mus. In this case
$\Phi_{1,5}=\Phi_{1,2}$. Note that for this value of $p$ there are no
kinks in\ \mp. However, in this case\ \hsydr\ agrees exactly with the
mass $m= 2M \sin\big({\pi\over 18}\big)$
(where $M$ is given by\ \Mlambda) of one of the
scalar particles of\ \mp. In fact, this theory contains two particles
with the masses
$$
m_1={m\over 2 \cos\big({\pi\over 18}\big)}\, , \ \
m_2=m\ .
$$
The formula\ \hsydr\ gives
$$
{\bar  \lambda}=-i\ 0.0785556...  \ m_1^{{18\over 7}}\ ,
$$
which is in good agreement with the Thermodynamic Bethe 
Ansatz calculation in\ \mus,
$$
{\bar \lambda}_{num}=-i\ 	0.0785551...  \ m_1^{{18\over 7}}\ .
$$

2. The model\ \mpp\ with $(p,p')=(2,9)$ was
studied in\ \martins . The model
contains four particles with the masses
$$
m_1\, , \quad
m_2=2 m_1 \cos\big({7\pi\over 30}\big) , 
\quad  m_3=2 m_1 \cos\big({\pi\over 15}\big) \, ,\quad
m_4=4 m_1 \cos\big({\pi\over 10}\big) \cos\big({7\pi\over 30}\big)\ .
$$
Calculations based on the Truncated Conformal Space
method\ \Yur\ in this model
give\ \martins 
$$
{\bar \lambda}_{num}=-i\ 0.013065...\  m_1^{{10\over 3}}\ .
$$
Identifying $m=m_2$ we obtain from\ \hsydr
$$
{\bar \lambda}=-i\ 0.0130454...\ m_1^{{10\over 3}}\ .
$$

3. The case $(p,p')=(3,14)$ in\ \mpp\ was investigated in\ \takacs.
There are
six particles with the masses
$$
m_1=m_2\, ,\quad m_3={\sqrt2} m_1 \, ,\quad
m_4=m_5= 2 m\,\cos\big({\pi\over 12}\big) \, ,\quad
m_6=2 \sqrt2 \,m\, \cos\big({\pi\over 12}\big)\ .
$$
The relation
$$
{\bar \lambda}_{num}=-i\ 0.011833...\ m_1^{{24\over 7}}
$$
given in\ \takacs\ is in good agreement with\ \hsydr\ which gives
$$
{\bar \lambda}=-i\  0.01183265...\ m_1^{{24\over 7}}\  ,
$$
provided we identify $m=m_3$.

It is interesting to notice 
that in all the above examples the mass $m$ is
related to the mass $m_1$ of the lightest particle in\ \mpp\ as
\eqn\hdyrrt{m_1={m\over 2 \sin({2\pi\xi\over 3-3\xi})}\ .}
We believe that
this is a general relation for\ \mpp\ which holds as long as
$1/5<\xi<5/9$.

\vskip 0.2in

\newsec{Expectation values of primary fields
in the minimal models perturbed by
the operator $\Phi_{2,1}$}

\vskip 0.1in

As is well known, the minimal models ${\cal M}_{p/p'}$ admit yet another
integrable perturbation
\eqn\mptwoone{{\cal M}_{p/p'}+ {\hat \lambda}\int d^2 x\, \Phi_{2,1}(x)\ .}
The theory\ \mptwoone\ makes sense for $2p > p'$; this condition guarantees
that the operator $\Phi_{2,1}$ is relevant. For $3p>2p'$ the vacuum
structure of\ \mptwoone\ is expected to be very similar to that of
\mp, with $p'$ playing the role of $p$. Namely, if $p'$ is odd, 
i.e. the perturbation is odd with respect
to the symmetry\ \ztwo, the theory has ${p'-1\over 2}$ degenerate
ground states which
we denote $\mid 0_s \rangle$ with $s=1, 3, ... , p'-2$ if
${\hat\lambda} < 0$ and $\mid 0_s \rangle$ with $s=2, 4, ... , p'-1$ if 
${\hat\lambda} > 0$. If $p'$ is even  and ${\hat\lambda} > 0$ there are
$p'/ 2-1$ ground states $\mid 0_s \rangle;\ s= 2, 4, ... , p'-2$,
while if $p'$ is even and ${\hat\lambda} < 0$, there are $p'/ 2$
vacua $\mid 0_s \rangle;\  s= 1, 3, ... , p'-1$ (compare this with the
situation in\ \mp\ discussed in Sect.4). The excitations are the kinks
interpolating between these vacua (and possibly some neutral
particles). The natural modification of the
conjecture\ \bdlk\ in this case is
\eqn\evtwoone{
\langle 0_s \mid \Phi_{l,k} \mid 0_s \rangle =
{{\sin\big(\, {{\pi s }\over {p'}}  | p'l-p k|\,\big)}
\over {\sin\big({{\pi s  }\over {p'}}  (p'-p) \,\big)}}\
\biggl[\,
 {M\, \pi\,  \xi\,
\Gamma\big({2\xi\over  3\xi-3}\big)\over
2^{2\over 3}\,
\sqrt3\,  \Gamma\big({1\over 3}\big)\, \Gamma\big({\xi+1\over 3\xi-3}\big)}
\, \biggr]^{2\Delta_{l,k}}\ {\cal Q}_{2,1}\big((\xi+1)l-\xi k\big)\ ,}
where the function
\eqn\qtwoone{\eqalign{{\cal Q}_{2,1}&(\eta)={\rm exp}\biggl\{
\int_{0}^{\infty} {d t\over t}\ \bigg(\,
{  {\rm sinh}\big((\xi-1) t\big)\  {\rm sinh}\big( t (\eta-1)\big)\
{\rm sinh}\big( t (\eta+1)\big)
\over   {\rm sinh}\big(3(\xi-1) t\big)\, {\rm sinh}\big( (\xi+1) t\big)\,
{\rm sinh}( 2 \xi  t)
}\times\cr
&\Big( \cosh\big(3(\xi-1) t\big)+
\cosh\big((\xi-3) t\big)-
\cosh\big((3\xi-1) t\big)+\cosh\big((\xi+1) t\big)+1 \Big)
-\cr
&\ \ \ \ \ \ \ \ \ \ \ \ \ \ \ \ \ \ \ \ \ \ \ \ \ \ \ \ \ \ \ \ \
{ (\eta^2-1 )\over 2  \xi (\xi+1)}\,  e^{-4t}\, \bigg)\biggr\}  }}
is obtained from\ \bdlk\ by the substitution $\xi \to -1-\xi$, and $M$
is the lightest kink mass. It is
remarkable that the function ${\cal Q}_{2,1}(\eta)$ coincides with
${\cal Q}_{1,5}(\eta)$ in\ \philksw\ continued to the domain $\xi >
2$.
For $(l,k)=(2,1)$\ \qtwoone\ gives
\eqn\hsydre{\eqalign{\langle\,\Phi_{2,1}\, \rangle =&(-1)^{s-1}\ 
{2^{2-{7\over 2\xi}}\, \xi\, \pi\,
\Gamma\big({3\xi-1\over 4\xi}\big)\,
\sin\big({\pi (\xi+1)\over 3\xi-3}\big)
\over
3\sqrt3\, (\xi-1)\,  \Gamma\big({\xi+1\over 4\xi}\big)\,
\sin\big({2\pi\, \xi\over 3\xi-3}\big)}\
\sqrt{{\Gamma\big({1\over 2}-{1\over \xi}\big)\over
\Gamma\big({1\over 2}+{1\over \xi}\big)}}\times
\cr &
\biggl[\,  {\sqrt3\,  \Gamma\big({1\over 3}\big)
\Gamma\big({\xi+1\over 3\xi-3}\big)\over
2\pi\, \Gamma\big({2\xi\over 3\xi-3}\big)}\, \biggr]^{{3 \xi-3\over 2\xi}}\
M^{{ \xi+3\over 2\xi}}\ .}}
This formula together with the expression
\eqn\jshdt{
f_{2,1}=-{M^2 \ \sin\big({\pi (\xi+1)\over 3\xi-3}\big)\over 4\, \sqrt3
\, \sin\big({2\pi\, \xi\over 3\xi-3}\big)} } 
for the specific free energy of\ \mptwoone\ leads to the following
relation between $M$ and ${\hat\lambda}$ in\ \mptwoone
\eqn\jdhfr{
M={2^{{\xi-4\over 3\xi-3}}\, \sqrt3\,
\Gamma\big({1\over 3}\big)\, \Gamma\big({\xi+1\over 3\xi-3}\big) \over
\pi\, \Gamma\big({2\xi\over 3\xi-3}\big)}\
\biggl[\,  {\pi^2\, {\hat \lambda}^2\, \Gamma^2\big({3\xi-1\over 4\xi}\big)\,
\Gamma\big({1\over 2}-{1\over \xi}\big)\over
\Gamma^2\big({\xi+1\over 4\xi}\big)\,
\Gamma\big({1\over 2}+{1\over \xi}\big) }\, \biggr]^{{\xi\over 3\xi-3}}\ ,}
which reproduces the result of\ \fatt.

Some checks of\ \evtwoone\ can be made. For $(p,p')=(3,4)$ the model
\mptwoone\ is just the off-critical Ising field theory with zero
magnetic field and
$\Phi_{1,2}$ coincides with the order parameter $\sigma$.
In this case all expectation values\ \evtwoone\ agree
with known result\ \WuM,
\eqn\gsrew{\langle 0_1 \mid \,\sigma\,\mid 0_1 \rangle=
-\langle 0_3 \mid \,\sigma\,\mid 0_3 \rangle=
M^{{1\over 8}}\, 2^{{1\over 12}}\, e^{-{1\over 8}} \,
{\cal A}^{{3\over 2}}\, ,\ 
\ \ \ \ \ \ \langle 0_2 \mid \,\sigma\,\mid 0_2 \rangle=0\ ,}
where ${\cal A}=1.282427...$ is Glaisher's constant and $M=
-2\pi\, \hat{\lambda}$.

In the case $(p,p')=(4,5)$,\ \mptwoone\ describes 
the tricritical Ising model
with sub-leading magnetic perturbation. The theory has two degenerate
ground states $\mid 0_2 \rangle$ and $\mid 0_4 \rangle$ (we assume that
${\hat\lambda}> 0$) and the formula\ \evtwoone\ gives
\eqn\evtric{\eqalign{&\langle 0_4 \mid \,\Phi_{2,2}\,\mid 0_4 \rangle=
-1.79745...\  {\hat \lambda}^{1\over 15}\ , \quad 
\langle 0_2 \mid \,\Phi_{2,2}\,\mid 0_2 \rangle=
0.68656...\  {\hat \lambda}^{1\over 15}\ ,\cr
&\langle 0_4 \mid \,\Phi_{1,2}\,\mid 0_4 \rangle= 
2.04451...\  {\hat \lambda}^{8\over 45}\ , \quad 
\langle 0_2 \mid \,\Phi_{1,2}\,\mid 0_2 \rangle=
-0.78093...\  {\hat \lambda}^{8\over 45}\ .}}
In\ \magnol\ the following 
numerical estimates for the expectation values of
these fields were obtained
\eqn\evnum{\eqalign{
&\langle\,\Phi_{2,2}\, \rangle_{num}\simeq
-1.09\ {\hat \lambda}^{1\over 15}\ , \cr
&\langle\,\Phi_{1,2}\, \rangle_{num}\simeq
1.2\  {\hat \lambda}^{8\over 45}\ .}}
Direct comparison between\ \evtric\ and \evnum\ is problematic because it
is not clear from\ \magnol\ precisely which ground state is taken in
calculating the expectation 
values\ \evnum. The calculations in\ \magnol\ are
based on Truncated Conformal
Space method\ \Yur\ where one starts with the finite-size system with the
spatial coordinate compactified on a circle of circumference $L$; the
estimates are then obtained by extrapolating the finite-size results to
$L=\infty$. Let us note in this connection that in the finite-size
system the above two ground states $\mid 0_2 \rangle$ and $\mid 0_4
\rangle$ appear in the limit $L\to \infty$ as particular superpositions
of two asymptotically degenerate states (with the energy splitting
$\sim e^{-ML}$ where $M$ is the kink mass) which we denote $\mid I
\rangle$ and $\mid II \rangle$ (the first being the true ground state at
finite $L$). Simple calculation which takes into account the known kink
picture of the excitations in this theory (see\ \smir,\ \zamole) gives
the following
exact relation between these states at $L=\infty$
\eqn\jsdfer{\eqalign{
&\mid I \rangle ={1\over 2\cos\big({\pi\over 10} \big) }\,
\mid 0_4 \rangle +{\cos\big({\pi\over 5}\big)\over 
\cos\big({\pi\over 10} \big) }\, \mid 0_2 \rangle\ ,\cr
&\mid II \rangle ={2\cos\big({\pi\over 10} \big)\over
\sqrt5 }\,
\mid 0_4 \rangle -{\cos\big({\pi\over 10}\big)\over
\sqrt5 \cos\big({\pi\over 5} \big) }\, \mid 0_2 \rangle\ .}}
It is easy to check using\ \evtwoone\ that
\eqn\evzero{\langle I \mid \Phi_{2,2} \mid I \rangle =
\langle I \mid \Phi_{1,2}
\mid I \rangle = 0\ .}
This result is not at all surprising as at finite $L$ the ground state
$\mid I \rangle$ is obtained by perturbing the conformal ground state
(i.e. the primary state $\mid \Phi_{1,1} \rangle$) by the operator
$\Phi_{2,1}$. Therefore at any finite $L$ the state $\mid I \rangle$ is
a superposition of the states out of the conformal families
$\big[\, \Phi_{1,1}\, \big],
\big[\,\Phi_{2,1}\, \big]$ and $\big[\, 
\Phi_{3,1}\, \big]$. By the conformal fusion
rules all these states produce vanishing expectation values of both 
$\Phi_{1,2}$ and $\Phi_{2,2}$. In fact, the result\ \evzero\ can be
considered as a nontrivial consistency check of our conjecture
\evtwoone. On the other hand, the expectation values
\eqn\evtwo{\eqalign{
&\langle II \mid \Phi_{2,2} \mid II \rangle =-\langle II \mid
\Phi_{2,2} \mid I \rangle=
1.11089...\ {\hat \lambda}^{{1\over 15}}\ ,\cr
&\langle II \mid \Phi_{1,2} \mid II \rangle =
\langle II \mid \Phi_{1,2} \mid I \rangle= 
1.26358...\ {\hat \lambda}^{8\over 45}\  ,}}
do not vanish and actually the numerical results\ \evnum\ appear to be 
reasonably close to\ $\langle II \mid \Phi_{2,2} \mid I \rangle$ and
$\langle II \mid \Phi_{1,2} \mid I \rangle$.
We believe therefore that it is these 
expectation values
that are quoted in\ \magnol.

\centerline{}

\centerline{\bf Acknowledgments}
 
\vskip0.5cm

S.L. is grateful to the Department of Physics and 
Astronomy, Rutgers University for the hospitality.
The work of S.L.
is supported in part by NSF grant.
Part of this work was done during A.Z.'s visit at 
the Laboratoire de Physique Math\'ematique, Universit\'e de
Montpelli\'er II, and the hospitality extended to him is
gratefully acknowledged.
Research of A.Z.
is supported in part by DOE grant \#DE-FG05-90ER40559.

S.L., A.Z. and Al.Z. are grateful to the organizers and participants
of the research program
``Quantum Field Theory in Low Dimensions: From
Condensed Matter to Particle Physics''
at the Institute for Theoretical Physics
at Santa Barbara (NSF grant No. PHY94-07194),
where parts of this work were done, for their
kind hospitality. 

\listrefs

\end